\documentclass[final,authoryear,5p,times,twocolumn]{elsarticle}

\usepackage{amssymb}

\begin{document}

\begin{frontmatter}



\title{Implementation of a Parallel Tree Method on a GPU}


\author[1]{Naohito Nakasato}
\address[1]{
Department of Computer Science and Engineering\\
University of Aizu\\
Aizu-Wakamatsu, Fukushima 965-8580, Japan\\
Email: nakasato@u-aizu.ac.jp
}

\begin{abstract}
The $k$d-tree is a fundamental tool in computer science. 
Among other applications, the application of $k$d-tree search (by the
tree method) to the fast evaluation of particle interactions and
neighbor search is highly important, since the computational
complexity of these problems is reduced from $O(N^2)$ for a brute
force method to $O(N \log N)$ for the tree method, where $N$ is
the number of particles. 
In this paper, we present a parallel
implementation of the tree method running on a graphics processing
unit (GPU). We present a detailed description of how we have
implemented the tree method on a Cypress GPU. 
An optimization that we found important is localized particle ordering to
effectively utilize cache memory. 
We present a number of test results
and performance measurements. 
Our results show that the execution of the tree traversal 
in a force calculation on a GPU is practical and efficient.
\end{abstract}


\end{frontmatter}


\section{Introduction}
A technique for gravitational many-body simulations is a
fundamental tool in astrophysical simulations because the
gravitational force drives structure formation in the universe.
The length scales that arise in structure formation range from
less than 1 cm for the aggregation of dust to more than $10^{24}$
cm for the formation of cosmological structures. 
At all scales, gravity is a key physical process for the understanding of
structure formation. 
The reason behind this is the long-range nature of gravity.

Suppose we simulate structure formation with $N$ particles. 
The flow of the many-body simulation is as follows. 
First, we calculate the mutual gravitational forces between the $N$ particles, 
then integrate the orbits for the $N$ particles, and repeat this process as necessary. 
Although it is simple in principle, the force calculation is a challenging task from the
point of view of computer science. 
A simple, exact method for the force calculation requires $O(N^2)$ computational complexity,
which is prohibitively compute-intensive for large $N$. 
An exact force calculation is necessary in some types of simulations, such
as few-body problems, the numerical integration of planets
orbiting around a star (e.g., the Solar System), and the evolution
of dense star clusters. 
For simulations that do not require exact
forces, however, several approximation techniques have been
proposed \citep{Hockney_1981, Barnes_1986, Greengard_1987}. The
particle--mesh/particle--particle--mesh method
\citep{Hockney_1981} and the tree method \citep{Barnes_1986}
reduce the computational complexity of the force calculation to
$O(N \log N)$. 
The fast multipole method (FMM) reduces it further to $O(N)$ \citep{Greengard_1987}. 
Of these methods, the tree method has been used extensively in astrophysical simulations,
since its adaptive nature is essential for dealing with clumpy
structure in the universe \citep[e.g.,][]{Bouchet_1988}.

Despite the $O(N \log N)$ complexity, computational optimization
of the tree method by techniques such as vectorization and
parallelization is necessary to accommodate demands for
simulations with larger and larger $N$. 
\cite{Hernquist_1990}, \cite{Makino_1990}, and \cite{Barnes_1990} have reported various
techniques to vectorize the force calculation with the tree method. 
\cite{Warren_1992}, \cite{Dubinski_1996}, and
\cite{Yahagi_1999} have reported a parallel tree method for
massively parallel processors (MPPs). 
In a recent publication \citep{Springel_2005}, a simulation of large-scale structure
formation in the universe with more than ten billion particles,
using a parallel tree code running on an MPP, has been reported.
Another computational technique to speed up the tree method
utilizes the GRAPE special-purpose computer \citep{Sugimoto_1990,Makino_1998}. 
Using a combination of vectorization techniques for the tree method, 
the tree method can be executed efficiently on a GRAPE system \citep{Makino_1991}.

Cosmological simulations are a ``grand challenge" problem. 
The Gordon Bell prizes have been awarded many times for cosmological
simulations \citep{Warren_1992a, Fukushige_1996, Warren_1997,
Warren_1998, Kawai_1999, Hamada_2009}. 
In those simulations, both parallel tree codes \citep{Warren_1992a, Warren_1997, Warren_1998}
and a tree code running on a GRAPE system \citep{Fukushige_1996, Kawai_1999}
and a graphics processing unit (GPU) \citep{Hamada_2009}
were used to perform cosmological simulations.

In the present paper, we describe our implementation of the tree method on a GPU. 
The rise of the GPU forces us to rethink our way of doing parallel computing, 
since the performance of recent GPUs has reached the impressive level of $> 1$ Tflops. 
Acceleration techniques for many-body simulations with a GPU have already been
reported \citep[e.g.,][]{Nyland_2007, Portegies_2007, Hamada_2007,Belleman_2008}; 
however, these techniques have implemented an
exact, brute force method with $O(N^2)$ complexity. 
It is apparent, however, that for applications that do not require exact
forces, it is possible to do much more efficient computation by
the tree method. 
We have directly implemented the tree method on a GPU so that we can enjoy the speed of an $O(N \log N)$ algorithm. 
For small $N < 30,000$, the brute force method on a GPU is faster than
the tree method owing to extra work concerning the tree data structure. 
However, our results show that the tree method significantly outperforms the brute force method on a GPU 
for $N \gg 10,000$, which is the standard size in current astrophysical
simulations is. 
Our code is simple and easy to extend to other numerical algorithms 
that require a neighbor list or a short-range force, 
such as algorithms for the smoothed particle hydrodynamics (SPH) method 
\citep{Gingold_1977, Lucy_1977}.

\section{GPU architecture}
In this section, we briefly summarize the architecture of the
Cypress GPU that we used in the present work (most of the
information is taken from \cite{AMD_2010}).

\subsection{Cypress architecture}
The Cypress GPU, from AMD, is the company's latest GPU and has
many enhancements for general-purpose computing. 
It has 1600 arithmetic units in total. 
Each arithmetic unit is capable of executing single-precision floating-point 
fused multiply--add (FMA) operation. 
Five arithmetic units make up a five-way
very-long-instruction-word (VLIW) unit called a stream core (SC).
Therefore, one Cypress processor has 320 SCs. 
One SC can execute a several combinations of operations such as
(1) five 32-bit integer operations, 
(2) five single-precision FMA operations, 
(3) four single-precision FMA operations with one transcendental operation, 
(4) two double-precision add operations, 
or (5) one double-precision FMA operations. 
Each SC has a register file of 1024 words, where one word is 128 bits long (four single-precision
words or two double-precision words). 
A group of 16 SCs make up a unit called a compute unit. 
At the top level of the GPU, there are 20 compute units, a controller unit called an ultra-threaded
dispatch processor, and other units such as units for graphics
processing, memory controllers, and DMA engines.

All SCs in the compute unit work in a
single-instruction-multiple-thread (SIMT) mode, i.e., 16 SCs
execute the same instructions for four clock cycles to accommodate
the latency of the arithmetic units. 
That is, we have 64 threads proceeding as a wavefront on the Cypress GPU. 
At the time of writing, the fastest Cypress processor runs at 850 MHz and offers
a peak performance of $1600 \times 2 \times 850 \times 10^6 = 2.72$ Tflop/s in single-precision operations. 
With double-precision operations, we have $320 \times 2 \times 850 \times 10^6 = 544$ Gflop/s.

\begin{figure*}[t]
\begin{center}
\includegraphics[width=13cm]{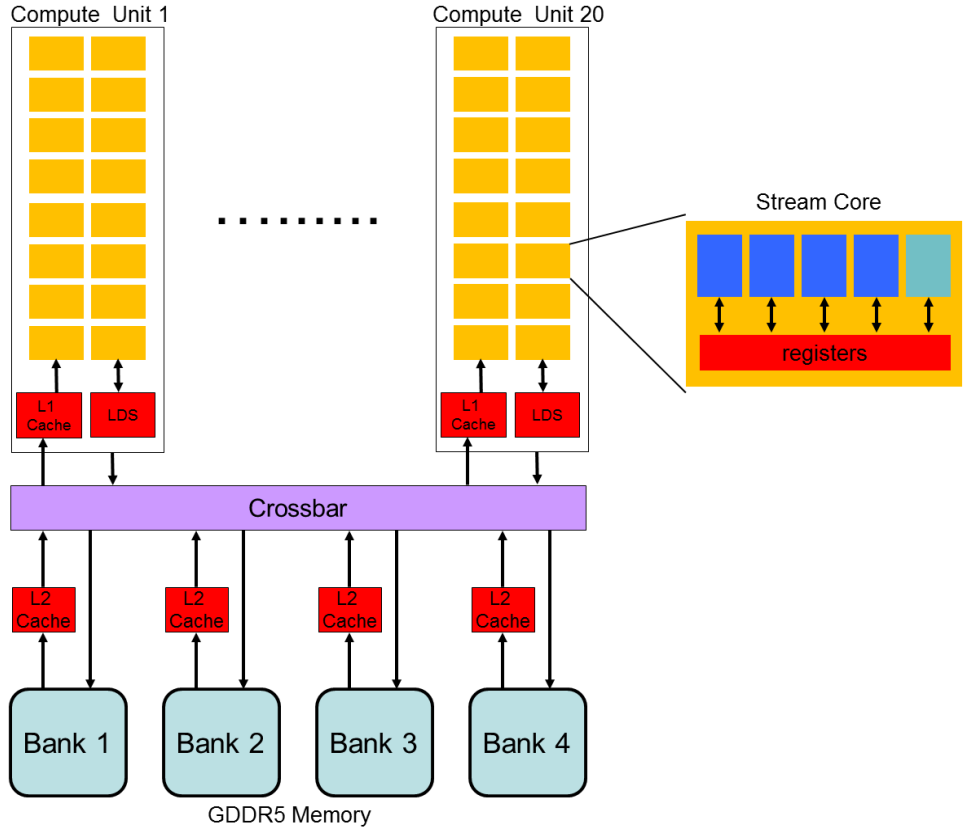}
\caption{ Block diagram of the Cypress GPU, with emphasis on the
memory system. } \label{Cypress}
\end{center}
\end{figure*}

The external memory attached to the Cypress consists of 1 GB of GDDR5 memory with a 256 bit bus. 
It has a data clock rate of 4.8
GHz and offers a bandwidth of 153.6 GB/s. This external memory is
accessed through four banks, as shown in Figure \ref{Cypress}. In
each bank, there is a second-level read cache (L2 cache). The
total size of the second-level cache is 512 KB, i.e., $4 \times
128$ KB. Twenty compute units and memory controllers are
interconnected through a crossbar. Each compute unit has a
first-level read cache (L1 cache) and a local data share (LDS), as
depicted in Figure \ref{Cypress}. 
The sizes of the L1 cache and
LDS are 8 KB and 32 KB, respectively. The L1 cache can fetch data
at 54.4 GB/s when the cache is hit; namely, the aggregate
bandwidth of the L1 cache on the Cypress GPU is 54.4 GB/s $\times$ 20 = 1.088 TB/s. 
This high memory bandwidth is a notable feature
of this GPU.
As we shall describe in the following section, taking
advantage of the hardware-managed cache is critical to obtaining
high performance on the Cypress GPU.

\subsection{Programming the Cypress GPU}
In the present work, we programmed the Cypress GPU using an
assembly-like language called IL (Intermediate Language). IL is
like a virtual instruction set for GPUs from AMD. With IL, we have
full control of every VLIW instruction. The programming model
supported by IL is a single-instruction-and-multiple-data (SIMD)
model at the level of the SC. In this programming model, 
a sequence of instructions generated from an IL program is executed
on all SCs simultaneously with different input data.

A block of code written in IL is called a compute kernel. 
The device driver for a GPU compiles IL instructions into the
corresponding machine code when we load a kernel written in IL. 
In a compute kernel, we explicitly declare what type of variable the
input data is. In the main body of the IL code, we write
arithmetic operations on the input data. Logically, each SC is
implicitly assigned data that is different from that for every
other SC. In the case of a simple compute kernel, the SC operates
on the assigned data. Operations such as this, as arise in pure
stream computing, seem to work with the highest efficiency. 
In a complex compute kernel, which we explore in the present work, each
SC not only operates on the assigned data but also explicitly
loads random data that might be assigned to another SC. 
To accomplish a random access to external memory, we explicitly
calculate the address of the data in the compute kernel.

The ATI Stream software development kit (SDK) for the Cypress GPU
also supports OpenCL, which is a standard API with an extended C
language (hereafter referred to as C for OpenCL) for writing a
compute kernel. In this work, we also present a compute kernel
written in C for OpenCL (see \ref{app} for the code). 
We believe that it is instructive to present our algorithm in C for OpenCL
and that this makes the algorithm easy to understand. 
Both programming methods (using IL and using C for OpenCL) have pros
and cons. 
With IL, we have the advantage of full control of the
VLIW instructions, but a compute kernel written in IL is somewhat
cumbersome. 
On the other hand, it is is easier to start writing a compute kernel
in C for OpenCL, but optimization for any particular GPU architecture
is not straightforward. 
An advantage of programming with OpenCL is
that we can use OpenCL to program a general-purpose many-core CPU.
In the following section, we compare implementations of the tree
method on a GPU based on compute kernels written in IL and in C for OpenCL. 
We also compare the performance of a compute kernel
written in C for OpenCL on a GPU and on a CPU.

\section{Bare performance of brute force method on a GPU}
So far, we have developed around a dozen kernels in IL that we use
for astrophysical many-body simulations. In this section, we
report the performance of our implementation of a brute force
method for computing gravitational forces. 
This code served as a basis for us to implement a more sophisticated algorithm later.

To be precise, we have implemented a set of conventional equations
expressed as
\begin{eqnarray}
p_i &=& \sum_{j=1,j \ne i}^{N} p(\mbox{\boldmath{$x$}}_i, \mbox{\boldmath{$x$}}_j, m_j) =
         \sum_{j=1,j \ne i}^{N} \frac{m_j}
                {(|\mbox{\boldmath{$x$}}_i - \mbox{\boldmath{$x$}}_j|^2 + \epsilon^2)^{1/2}}, \nonumber \\
\mbox{\boldmath{$a$}}_{i} &=& \sum_{j=1,j \ne i}^{N} \mbox{\boldmath{$f$}}(\mbox{\boldmath{$x$}}_i, \mbox{\boldmath{$x$}}_j, m_j) = \sum_{j=1,j \ne i}^{N} \frac{m_j (\mbox{\boldmath{$x$}}_i - \mbox{\boldmath{$x$}}_j)}
                {(|\mbox{\boldmath{$x$}}_i - \mbox{\boldmath{$x$}}_j|^2 + \epsilon^2)^{3/2}}, \nonumber \\
\label{gravity}
\end{eqnarray}
where $\mbox{\boldmath{$a$}}_{i}$ and $p_i$ are the force vector
and potential for a particle $i$, and $\mbox{\boldmath{$x$}}_i$,
$m_i$, $\epsilon$ are the position of the particle, its mass, and
a parameter that prevents division by zero, respectively. 
We can solve these equations by two nested loops on a general-purpose CPU. 
In the inner loop, we simultaneously evaluate the functions
$p$ and $\mbox{\boldmath{$f$}}$, and require 22 arithmetic
operations, which include one square root and one division, to
compute the interaction between particles $i$ and $j$. 
Since previous authors, starting from \cite{Warren_1997}, 
have used a conventional operation count for the evaluation of
$\mbox{\boldmath{$f$}}_{i}$ and $p_i$, we have adopted a
conventional count of 38 throughout this paper.

\begin{figure}[t]
\centering
\includegraphics[width=6cm,angle=-90]{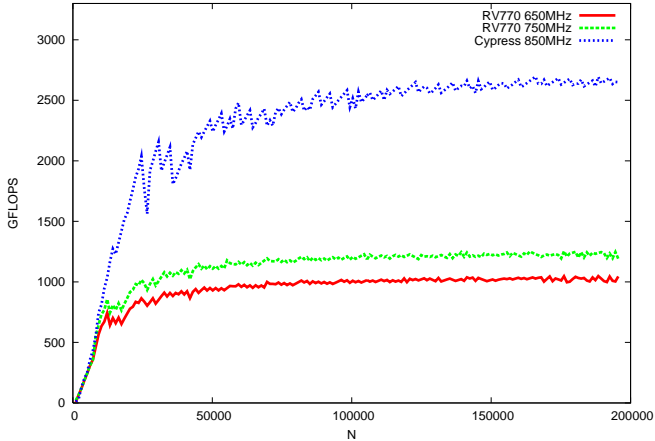}
\caption{Performance of the brute force method on various GPUs}
\label{bare}
\end{figure}

\cite{Elsen_2006} reported an implementation of a brute force
method for gravitational and other forces on an old GPU from
AMD/ATi. 
One of the main insights obtained was that a
loop-unrolling technique greatly enhanced the performance of the
code. 
We have followed Elsen et al.'s approach and tried several
different methods of loop unrolling. 
\cite{Fujiwara_2009} have
reported our optimization efforts for old GPUs. 
Here, we present a summary of our results. 
In Figure \ref{bare}, we plot the computing speed of our optimized IL kernel 
for computing Eq.(\ref{gravity}) as a function of $N$. 
We tested three GPU boards, namely RV770 GPUs running at 625 and 750 MHz 
and a Cypress GPU running at 850 MHz. 
The three systems had peak computing speeds in
single precision of 1.04, 1.2, and 2.72 Tflop/s, respectively. 
So far, we have obtained a maximum performance of $\sim$ 2.6 Tflop/s
on the Cypress GPU for $N > 150,000$. 
For $N = 195,584$, our optimized brute force method took roughly 0.5 s on the Cypress GPU. 
As far as we know, the performance that we obtained is the
fastest ever with one GPU chip.

Even with the massive computing power available on such GPUs,
however, we cannot escape from a computational complexity of $O(N^2)$. 
Therefore, if we need to do an astrophysical many-body simulation for large $N$, 
we need a smart algorithm to do the job,
since the recent standard for $N$ in astrophysical simulations is
at least $100,000$ for complex simulations with baryon physics and
$1,000,000$ for simple many-body simulations.

\section{Tree method on a GPU}
\subsection{Tree method}
The tree method \citep{Barnes_1986} is a special case of the general $k$d-tree algorithm. 
This method has been optimized to efficiently calculate the mutual forces between particles, 
and reduces the computational complexity of the force calculation from
$O(N^2)$ for the brute force method to $O(N \log N)$. 
A trick used is that instead of computing the exact force by a brute force
method, it approximates the force from distant particles using a multipole expansion. 
It is apparent that there is a trade-off
between the approximation error and the way in which we replace a
group of distant particles by a multipole expansion. 
A tree structure that contains all particles is used to judge this
trade-off efficiently.

The force calculation in the tree method is executed in two
steps: (1) a tree construction and (2) the force calculation. 
In the tree construction, we divide a cube that encloses all of the
particles into eight equal sub-cells. 
The first cell is the root of a tree that we construct; it is called the root cell. 
Then, each sub-cell is recursively subdivided in the same say until each cell
contains zero or one particle. 
As the result of this procedure, we obtain an oct-tree.

\begin{figure}
\centering
\begin{verbatim}
procedure treewalk(i, cell)
  if cell has only one particle
    force += f(i, cell)
  else
    if cell is far enough from i
      force += f_multipole(i, cell)
    else
      for i = 0, 7
        if cell->subcell[i] exists
          treewalk(i, cell->subcell[i])
\end{verbatim}
\caption{Pseudocode for the force calculation by traversing the tree} \label{treewalk}
\end{figure}

In the force calculation, we traverse the tree to judge whether
we should replace a distant cell that contains a group of
particles that are geometrically close together with the multipole
expansion of those particles. 
If we do not replace the cell, we then traverse sub-cells of the distant cell. 
If we do replace the cell, we calculate a particle--cell interaction. 
When we encounter a particle, we immediately calculate a particle--particle
interaction. 
Given a particle with index \verb|i| which we want to
compute the force acting on, this procedure is expressed as
pseudocode in Figure \ref{treewalk}. 
Note that \verb|subcell[]| is a pointers to its own sub-cells. 
In this pseudocode, \verb|f| is a function that computes the particle--particle interaction, 
and \verb|f_multipole| is a function that computes the particle--cell
interaction. 
In the work described in this paper, since we
considered only the monopole moment of a cell, both functions were
expressed exactly as in Eq. (\ref{gravity}). 
In principle, we can use any high-order moment in the particle--cell interaction.

We follow this procedure starting from the root cell, with the
following condition that tests whether a cell is far enough away.
Let the distance between the particle and the cell be $d$. 
The cell is well separated from the particle if $l/d < \theta$, where
$l$ is the size of the cell and $\theta$ is a parameter that
controls the trade-off. 
Since the smaller $l/d$ is, the more distant the cell is from the particle, 
this condition (called the opening condition) tests geometrically whether the cell is far
from the particle. 
This recursive force-calculation procedure is 
almost the same as in the original algorithm of
\cite{Barnes_1986}.

An important feature of the tree method is that with tree
traversal, the force calculations for different particles are
completely independent of each other. 
Therefore, after we have completed the tree construction, 
the force calculation is a massively parallel problem. 
There are two possible ways to implement the tree method
on a GPU to take advantage of this feature.

\subsection{Tree method with GRAPE}
One way is a method proposed by \cite{Makino_1991}. This method
was proposed as a tree method for the special-purpose computer
GRAPE. A GRAPE system consists of a host computer and a GRAPE
board or boards. The host computer controls the GRAPE board. 
For a program running on the host, the GRAPE board acts like a
subroutine that calculates the gravitational forces for given
particles.

So, we need the following two steps to use a GRAPE system for a
force calculation using the tree method: (1) construction of an
interaction list on the host computer, and (2) the actual force
calculation on the GRAPE board. 
The interaction list is a list of
particles and distant cells that are supposed to interact with a
given particle. 
After the construction of interaction lists for
all particles is completed, we compute the force on each particle
by sending the interaction lists to the GRAPE board. 
These two steps are necessary because the GRAPE board does not have the
ability to traverse the tree. 
Many authors have used this method extensively. 
Three winners and a finalist of the Gordon Bell prize
have used a variant of this method with a different version of the
GRAPE system and a GPU \citep{Fukushige_1996, Kawai_1999,
Hamada_2009, Kawai_2006}. 
A drawback of this approach is that the
performance is limited by the speed of the host computer that is
responsible for the tree traversal. 
This possible bottleneck,
which is similar to Amdahl's law, might be critical without a
highly tuned implementation of \verb|treewalk()| running on the host. 
Furthermore, in all of the results presented by
\cite{Fukushige_1996}, \cite{Kawai_1999}, \cite{Kawai_2006}, and
\cite{Hamada_2009}, extra force evaluations by a factor of two
were required to obtain the best performance. 
Note that because of the extra force evaluations, the maximum error in the force that
these authors have reported was better than the error obtained
with the conventional tree method for a given $\theta$.

\subsection{General tree walk}
Another way to implement the tree method, which we have followed
in the present work, is to implement the whole procedure shown in
Figure \ref{treewalk} on a GPU. 
The advantage of this approach is that only the tree construction, 
which requires relatively little time, is executed on the host, so that we utilize the massive
computing power of the GPU as much as possible. 
More importantly, we can use our method in applications that require short-range
interaction forces \citep{Warren_1995}. 
This is because it is possible to implement a neighbor search algorithm as a general
tree-walk procedure of the kind shown in Figure \ref{general_treewalk}. 
Two procedures, \verb|proc_particle| and \verb|proc_cell|, 
are used to process the particle--particle and
particle--cell interactions, respectively. 
In addition, a function \verb|distance_test| is used to control the treatment of a distant cell. 
The calculation of the gravitational force is an application
of the general tree-walk procedure that has been very successful.

\begin{figure}
\centering
\begin{verbatim}
procedure general_treewalk(i, cell)
  if cell has only one particle
    proc_particle(i, cell)
  else
    if distance_test(i, cell) is true
      proc_cell(i, cell)
    else
      for i = 0, 7
        if cell->subcell[i] exists
          general_treewalk(i, cell->subcell[i])
\end{verbatim}
\caption{Pseudocode for a general tree-walk procedure.}
\label{general_treewalk}
\end{figure}

\subsection{Our GPU implementation}
In our implementation of the tree method on a Cypress GPU, we
first construct an tree on the host computer that controls the GPU. 
At this stage, there is no difference between our original
tree code and the newly developed code for the GPU.

\begin{figure*}[t]
\centering
\includegraphics[width=13cm]{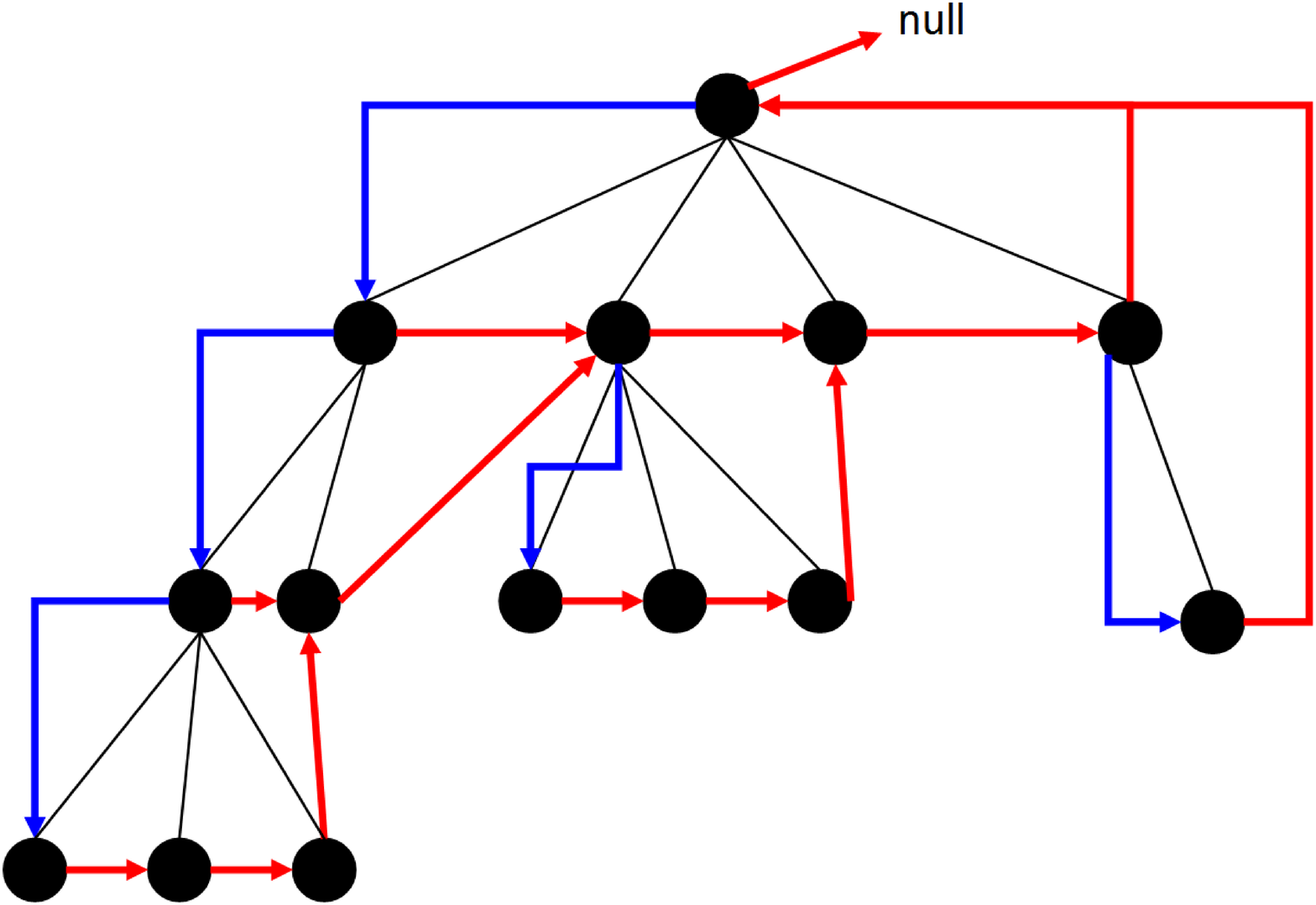}
\caption{A tree with ``more'' and ``next'' pointers, shown by blue
and red arrows, respectively.} \label{tree}
\end{figure*}

We need to take special care in implementing the tree-walk
procedure on the GPU. 
Currently, GPU architecture does not support
recursive procedures except when it is possible to fully expand a
recursion. 
Such a full expansion is possible only if the level of
the recursion is fixed, but in the tree method, it is impossible
to know how deep the recursion will be without performing the tree
traversal. 
So, we adopted a method proposed by \cite{Makino_1990}
that transforms a recursion in \verb|treewalk()| into an
iteration. 
A key feature is that for a given cell, we do not need
whole pointers (\verb|subcell[]|) to traverse the tree. 
We need only two pointers, to the cells that we will visit next when the
opening condition is true and when it is false, respectively.
These two pointers (hereafter called \verb|next[]| and
\verb|more[]|) are easily obtained by a breadth-first traversal of the tree. 
Figure \ref{tree} shows \verb|next[]| and \verb|more[]| schematically. 
Note that a cell that has sub-cells has both a
\verb|next[]| and a \verb|more[]| pointer, while a leaf cell 
(a particle in the present case) with no sub-cells has only a
\verb|next[]| pointer. 
An iterative form of \verb|treewalk()| with
these two pointers is shown in Figure \ref{treewalk_iterative}.
\label{NMp}

\begin{figure}
\centering
\begin{verbatim}
procedure treewalk_iterative(i)
  cell = the root cell
  while cell is not null
    if cell has only one particle
      force += f(i, cell)
      cell = cell->next
    else
      if cell is far enough from i
        force += f_multipole(i, cell)
        cell = cell->next
      else
        cell = cell->more
\end{verbatim}
\caption{Pseudocode for an iterative tree-walk procedure.}
\label{treewalk_iterative}
\end{figure}

We implemented the iterative procedure \verb|treewalk()| rather
directly in IL. The input data for this compute kernel is four arrays. 
The first contains the positions and masses of the
particles and cells. We pack a position and a mass into a vector
variable with four elements. Therefore, this array is an array of
four-element vectors. The mass of a cell equals the total mass of
the particles in the cell, and the position of the cell is at the
center of mass of the particles. The second and third arrays
contain the ``next'' and ``more'' pointers, respectively. 
Both of these are simple arrays. The fourth array contains the sizes of
the cells. The size of the cell is necessary for testing the
opening condition. See the description in Appendix \ref{app} for
the definitions of these arrays.

\begin{figure*}[t]
\centering
\includegraphics[width=13cm, angle=-90]{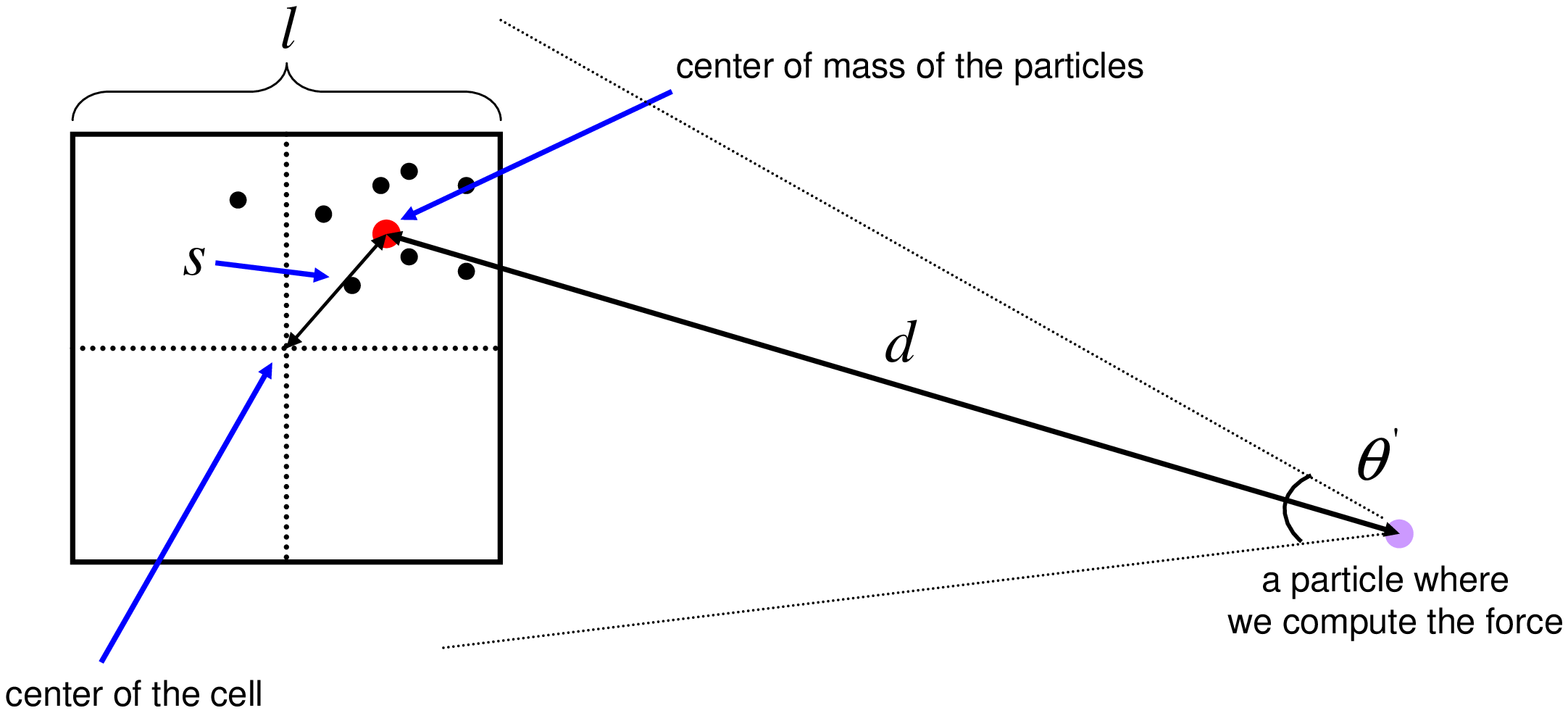}
\caption{Schematic view of a distant cell and a particle (shown by
a solid purple point). The black solid points are particles that
belong to the cell. The large red point is the center of mass of
the particles in the cell. } \label{cell-theta}
\end{figure*}

In the present work, we adopted the following modified opening
condition, expressed as
\begin{equation}
\frac{l}{\theta} + s < d,
\label{mac}
\end{equation}
where $s$ is the distance between the center of the cell and the
center of mass of the cell. 
The modified condition of Eq. (\ref{mac}) takes the particle distribution in the cell into
account through $s$ since if particles gather at a corner of a
cell, the effective size of the cell becomes larger. 
In Figure \ref{cell-theta}, we present a schematic view of a distant cell
and a particle which we are trying to calculate the force acting on. 
In practice, we precomputed the square of the effective size
$S_{\rm effective}$ as
\begin{equation}
S_{\rm effective} = \left( \frac{l}{\theta} + s \right)^2,
\end{equation}
and sent $S_{\rm effective}$ instead of $l$ for each cell.
With $S_{\rm effective}$, we do not need to compute the square
root of $d$, and we simply compare $S_{\rm effective}$ and $d^2$
during the tree traversal.

In Figure \ref{treegpu}, we present an abstract version of our
compute kernel written in IL. In IL programming, each SC executes
the compute kernel with the assigned data in parallel. In this
code, \verb|own| represents the specific cell assigned to each SC;
\verb|=|, \verb|load|, and \verb|->| are not real IL instructions
or operations but conventional symbols used here for the purpose
of explanation. We have omitted the calculation of the load
addresses for the arrays since it is too lengthy to show in
detail. In addition, the particle--particle and particle--cell
interaction codes have been omitted because they simply compute
the functions $\mbox{\boldmath{$f$}}$ and $p$ in Eq.
(\ref{gravity}). In \ref{app}, we present a working compute kernel
written in C for OpenCL. We present a performance comparison
between the IL and OpenCL implementations in the next section.

\begin{figure}
\centering
\begin{verbatim}
...declaration of I/O arrays and constants...
...initialize variables for accumulation...

xi = load own->x
yi = load own->y
zi = load own->z

cell = root
whileloop
  break if cell is null

  xj = load cell->x
  yj = load cell->y
  zj = load cell->z
  mj = load cell->m
  s_eff = load cell->s_eff

  dx = xj - xi
  dy = yj - yi
  dz = zj - zi
  r2 = dx*dx + dy*dy + dz*dz

  if cell is a particle
    ...compute particle-particle interaction...
    cell = load next
  else
    if r2 > s_eff
      ...compute particle-cell interaction...
      cell = load cell->next
    else
      cell = load cell->more
  endif
endloop
\end{verbatim}
\caption{Abstract IL code for our compute kernel that executes the
iterative tree walk.} \label{treegpu}
\end{figure}

With the compute kernel shown, the flow of our tree method on the Cypress GPU is as follows.
\begin{enumerate}
\item Construct a tree ({\bf host}). 
\item Compute the total mass, the center of mass, and the effective size of each cell ({\bf host}). 
\item Compute the ``next" and ``more" pointers ({\bf host}). 
\item Send the input data to the GPU ({\bf host}). 
\item Iterative tree walk associated with the force calculation for each particle ({\bf GPU}). 
\item Receive the force for each particle from the GPU ({\bf host}).
\end{enumerate}
We have indicated whether the corresponding part is executed on
the host or the GPU in bold text at the end of each step.

\section{Tests and optimization}
\label{btest}

\begin{table}
\renewcommand{\arraystretch}{1.3}
\caption{Our test system}
\label{conf}
\begin{center}
\begin{tabular}{c|c}
\hline
CPU & Intel Xeon E5520 $\times$ 2 \\
Memory & DDR3 800 1GB $\times$ 6 \\
GPU & Radeon 5870 memory 1GB \\
OS  & Ubuntu 9.10 (64 bit) \\
Driver & Catalyst 10.8 (fglrx 8.76.7 [Aug  3 2010]) \\
SDK & ATI Stream SDK 2.2 \\
\hline
\end{tabular}
\end{center}
\end{table}

Here, we describe the results of some basic tests to show that our
code worked correctly, and to obtain some performance characteristics. 
We used the configuration shown in Table
\ref{conf} for all results presented in this paper. 
In the basic tests, we used a set of particles randomly distributed in a sphere.

First, Table \ref{bench1} shows how the computing time depends on $N$. 
Each value of computing time was obtained by averaging the
results of 20 runs. 
In this test, we set $\theta = 0.6$. 
$T_{\rm total}$ and $T_{\rm construction}$ are the total time required for
the force calculation and the time spent on the construction of
the tree, respectively. 
Roughly, the tree construction took 20--28\% of $T_{\rm total}$. 
For all values of $N$ used, we checked that there was effectively no error in the force computed
by the GPU. 
\footnote{Here, the ``error'' is not the error due to
the approximations in the force calculations.} 
All operations on the GPU were done with single-precision, and we observed that the
error was comparable to the machine epsilon, $\sim 10^{-6}$. 
We believe that the error originates from a difference in the
implementations of the inverse of the square root on the host and
on the GPU. We consider that this is not at all significant for
our purpose of astrophysical many-body simulations.

\begin{table}
\renewcommand{\arraystretch}{1.3}
\caption{Dependence of computing speed on $N$: no sorting}
\label{bench1}
\begin{center}
\begin{tabular}{c|c|c}
\hline
$N$ & $T_{\rm total}$ (s) & $T_{\rm construction}$ (s) \\
\hline
50K  & $3.95 \times 10^{-2}$ & $1.1 \times 10^{-3}$ \\
100K & $9.65 \times 10^{-2}$ & $2.5 \times 10^{-2}$ \\
200K & $2.34 \times 10^{-1}$ & $6.1 \times 10^{-2}$ \\
400K & $5.83 \times 10^{-1}$ & $1.5 \times 10^{-1}$ \\
800K & $1.36 \times 10^{0}$  & $3.8 \times 10^{-1}$ \\
\hline
\end{tabular}
\end{center}
\end{table}

Regarding computing speed, randomly distributed particles are the
most severe test because two successive particles in the input
data have a very high chance of being at different positions. 
By the nature of the tree method, if two particles are close to each
other, those particles are expected to be in the same cell and to
interact with a similar list of particles and distant cells. 
This means that if two successive particles in the input data are
geometrically close, the tree walk for the second particle almost
certainly takes less time owing to a higher cache-hit rate. 
To accomplish such a situation, we can sort the particles to ensure
that successive particles are as close as possible together.
Fortunately, such sorting is easily available with the tree method
by traversing the tree in depth-first order. In the course of the
traversal, we add each particle encountered at a leaf node to a
list. After the tree traversal, we can use the list obtained to
shuffle the particles so that the order of particles is nearly the
desired order. 
This ordering of particles is called the Morton ordering. 
With this preprocessing, the speed of our method was
altered as shown in Table \ref{bench2}. 
Note that the time in Table \ref{bench2} does not contain the time required for the
preprocessing. This is adequate, since in astrophysical many-body
simulations, the tree structure is repeatedly constructed at each
time step so that we can automatically obtain this sorting for
free. We observed that $T_{\rm total}$ obtained with the Morton
ordering was faster by a factor of 1.5--2.2, depending on $N$,
than without the preprocessing. Moreover, $T_{\rm construction}$
also decreased in all cases owing to better cache usage on the
host. With the Morton ordering, the tree construction took roughly
14--27\% of $T_{\rm total}$.

\begin{table}
\renewcommand{\arraystretch}{1.3}
\caption{Dependence of computing speed on $N$: particles sorted in
Morton order} \label{bench2}
\begin{center}
\begin{tabular}{c|c|c}
\hline
$N$ & $T_{\rm total}$ (s) & $T_{\rm construction}$ (s) \\
\hline
50K  & $3.00 \times 10^{-2} $ & $9.1 \times 10^{-3} $ \\
100K & $6.08 \times 10^{-2} $ & $1.8 \times 10^{-2} $ \\
200K & $1.27 \times 10^{-1} $ & $3.9 \times 10^{-2} $ \\
400K & $2.65 \times 10^{-1} $ & $8.0 \times 10^{-2} $ \\
800K & $5.66 \times 10^{-1} $ & $1.6 \times 10^{-1} $ \\
\hline
\end{tabular}
\end{center}
\end{table}

The programming API for the Cypress GPU has a facility to report
the cache-hit rate for the GPU. In Table \ref{cache}, we show how
the cache-hit rate depends on $N$ and the ordering of the
particles. The results indicate that the performance of our method
is significantly affected by the ordering of the particles. In the
tests described in the following, we always used preprocessing.
Note that we could have obtained even better results if we had
sorted the particles in the Peano--Hilbert order, which has been
reported to be the optimal order for locality of data access, and
is used by some tree codes \citep[e.g.,][]{Warren_1993}.

\begin{table}
\renewcommand{\arraystretch}{1.3}
\caption{Dependence of cache-hit rate on $N$ for different
orderings of the particles} \label{cache}
\begin{center}
\begin{tabular}{c|c|c}
\hline
$N$ & No sorting (\%) & Morton ordering (\%) \\
\hline
50K  & 75 & 93 \\
100K & 63 & 91 \\
200K & 55 & 87 \\
400K & 48 & 85 \\
800K & 43 & 80 \\
\hline
\end{tabular}
\end{center}
\end{table}

\begin{figure}[t]
\centering
\includegraphics[width=6cm, angle=-90]{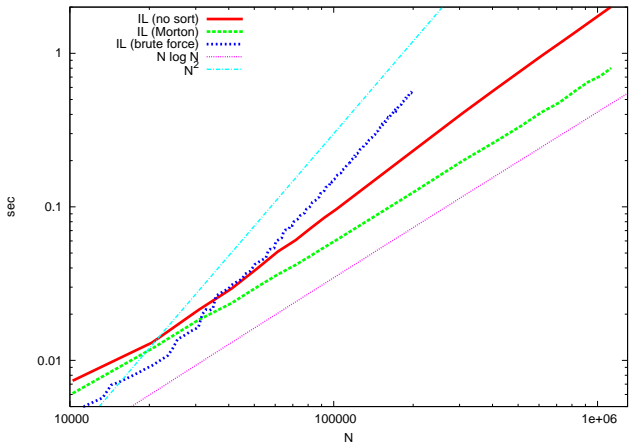}
\caption{Comparison between the tree method on a GPU and the
brute force method on a GPU. $T_{\rm total}$ as a function of $N$
is plotted for three cases. The $N^2$ and $N \log N$ scaling lines
are also plotted for reference. } \label{time}
\end{figure}

In Figure \ref{time}, we present $T_{\rm total}$ as a function of
$N$ for three cases: the tree method with Morton ordering, the
tree method without sorting, and the brute force method. 
Except for $N < 30,000$, the tree method with Morton ordering ($\theta = 0.6$) 
outperforms the brute force method on the GPU.

\begin{figure}[t]
\centering
\includegraphics[width=6cm, angle=-90]{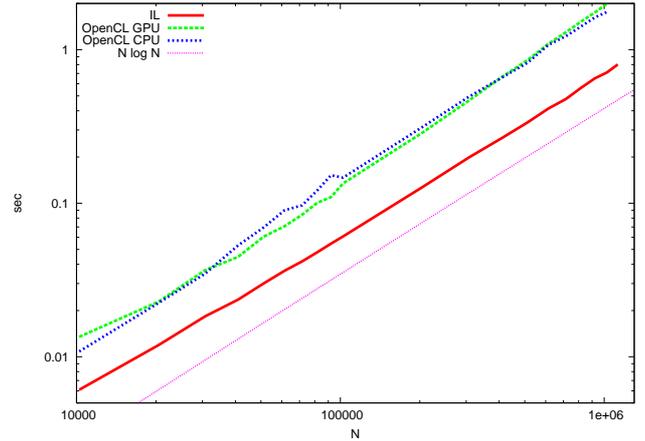}
\caption{Comparison between the tree method using a kernel
written in IL, the tree method using a kernel written in OpenCL
on a GPU, and the tree method using a kernel written in OpenCL
on a CPU. The $N \log N$ scaling line is also plotted for
reference. } \label{time0}
\end{figure}

In Figure \ref{time0}, we compare the performance for the
following three cases: (a) a kernel written in IL running on a
Cypress GPU, (b) a kernel written in C for OpenCL running on a
Cypress GPU, and (c) a kernel written in C for OpenCL running on a
multicore CPU. Since our test system had eight physical (16
logical) cores, the OpenCL kernel ran on the CPU with 16 threads.
The last two cases show almost identical performance even though
the theoretical performance in single precision for the Cypress
GPU is $\sim$2 0 times faster. 
In fact, the tree method is not
compute-intensive but is limited by memory bandwidth, and hence
the effective performance of the compute kernel written in IL is
roughly $\sim$ 1\% of the theoretical performance in single-precision. 
On the other hand, the performance gap between the two
kernels written in IL and C for OpenCL is a factor of $\sim$2.5.
We believe that one of the main reasons is that our compute kernel
written in C for OpenCL is executed without using L1 cache. 
We will investigate further optimizations of the OpenCL kernel in future work.

Next, we examine how $T_{\rm total}$ depends on $\theta$, which
controls the error bound for the tree method. 
A larger $\theta$ means that more of the distant particles are replaced by a
multipole expansion. In other words, for a smaller $\theta$, we
need to perform a larger number of force calculations, and hence
the computation will take a longer time. At the same time, the
error due to the multipole expansion decreases. Practically, a
force calculation by the tree method with $\theta < 0.1$ is
reduced to almost the same level as a brute-force computation. 
In such a regime, effectively, we do not have any preference for the
tree method. 
In Table \ref{theta}, we show the dependence of
$T_{\rm total}$ and the cache-hit rate on $\theta$. In this test,
we used $N = 800$K particles. In all of the tests that we have
presented so far, a clear trend is that the computing time seems
to be determined solely by the cache-hit rate. Before we had done
these tests, we expected that branch operations would be a
bottleneck for the compute kernel. 
In reality, all that matters is the cache-hit rate.

\begin{table}
\renewcommand{\arraystretch}{1.3}
\caption{Dependence of $T_{\rm total}$ and the cache-hit rate on
$\theta$ for $N$ = 800K} \label{theta}
\begin{center}
\begin{tabular}{c|c|c}
\hline
$\theta$ & $T_{\rm total}$ (s) & Cache-hit rate (\%)\\
\hline
0.2 & $ 1.65 \times 10^{1} $ & 23 \\
0.3 & $ 5.24 \times 10^{0} $ & 36 \\
0.4 & $ 2.24 \times 10^{0} $ & 48 \\
0.5 & $ 1.14 \times 10^{0} $ & 59 \\
0.6 & $ 6.82 \times 10^{-1}$ & 68 \\
0.7 & $ 4.92 \times 10^{-1}$ & 75 \\
0.8 & $ 3.98 \times 10^{-1}$ & 80 \\
0.9 & $ 3.43 \times 10^{-1}$ & 80 \\
1.0 & $ 3.10 \times 10^{-1}$ & 82 \\
\hline
\end{tabular}
\end{center}
\end{table}

Finally, we measured the average error in the accelerations using
the following equation:
\begin{equation}
a_{\rm error} = \frac{1}{N} \sum_{i = 0}^{N-1}
\frac{|\mbox{\boldmath{$a$}}_{i}^{\rm tree} -  \mbox{\boldmath{$a$}}_{i}^{\rm direct}|}
{|\mbox{\boldmath{$a$}}_{i}^{\rm direct}|},
\end{equation}
where $\mbox{\boldmath{$a$}}_{i}^{\rm tree}$ and
$\mbox{\boldmath{$a$}}_{i}^{\rm direct}$ are the acceleration
forces obtained by the tree method on the GPU and by the brute
force method on the host computer, respectively. We use a similar
definition for the error in the potential, $p_{\rm error}$. For
this test, we used a realistic particle distribution that
represented a disk galaxy. 
We used GalactICS \citep{Kujiken_1995} to generate the particle distribution. 
The particle distribution had three components, namely a bulge, a disk, and a halo, with a
mass ratio of approximately 1:2:12. Tables \ref{ERR_a} and
\ref{ERR_p} present $a_{\rm error}$ and $p_{\rm error}$,
respectively, for several different values of $N$ and $\theta$.
Both $a_{\rm error}$ and $p_{\rm error}$ depend on $\theta$ as
expected. Except for $N$ = 500K, we have a smaller $a_{\rm error}$
for larger $N$. We found no systematic error in
$\mbox{\boldmath{$a$}}_{i}$ computed by the tree code on the GPU. 
However, it would be desirable to use double-precision
variables for accumulation of $\mbox{\boldmath{$a$}}_{i}$ to reduce $a_{\rm error}$ for large $N > 10^5$. 
There is only a negligible difference between the results computed by the compute
kernels written in IL and in C for OpenCL.

\begin{table*}
\renewcommand{\arraystretch}{1.3}
\caption{Dependence of the average acceleration error $a_{\rm
error}$ on $N$ and $\theta$. $N$ is a multiple of 1024.}
\label{ERR_a}
\begin{center}
\begin{tabular}{c|c|c|c|c|c}
\hline
$\theta$ & $N=10$K & $N=50$K & $N=100$K & $N=200$K & $N=500$K \\
\hline
0.2 & 2.93e-04 & 1.69e-04 & 1.48e-04 & 2.24e-04 & 5.73e-03 \\
0.3 & 6.37e-04 & 4.42e-04 & 3.98e-04 & 4.52e-04 & 5.90e-03 \\
0.4 & 1.23e-03 & 8.86e-04 & 8.16e-04 & 8.27e-04 & 6.20e-03 \\
0.5 & 2.04e-03 & 1.50e-03 & 1.41e-03 & 1.36e-03 & 6.64e-03 \\
0.6 & 3.15e-03 & 2.31e-03 & 2.20e-03 & 2.06e-03 & 7.20e-03 \\
0.7 & 4.39e-03 & 3.35e-03 & 3.18e-03 & 2.94e-03 & 7.91e-03 \\
0.8 & 5.94e-03 & 4.51e-03 & 4.33e-03 & 3.98e-03 & 8.77e-03 \\
0.9 & 7.85e-03 & 5.95e-03 & 5.69e-03 & 5.22e-03 & 9.86e-03 \\
1.0 & 9.95e-03 & 7.69e-03 & 7.34e-03 & 6.81e-03 & 1.13e-02 \\
\hline
\end{tabular}
\end{center}
\end{table*}

\begin{table*}
\renewcommand{\arraystretch}{1.3}
\caption{Dependence of the average potential error $p_{\rm error}$
on $N$ and $\theta$.} \label{ERR_p}
\begin{center}
\begin{tabular}{c|c|c|c|c|c}
\hline
$\theta$ & $N=10$K & $N=50$K & $N=100$K & $N=200$K & $N=500$K \\
\hline
0.2 & 4.46e-05 & 4.39e-05 & 5.37e-05 & 5.52e-05 & 4.64e-05  \\
0.3 & 9.87e-05 & 1.07e-04 & 1.41e-04 & 1.42e-04 & 1.02e-04  \\
0.4 & 1.84e-04 & 1.90e-04 & 2.70e-04 & 2.66e-04 & 1.80e-04  \\
0.5 & 2.98e-04 & 2.96e-04 & 4.30e-04 & 4.23e-04 & 2.88e-04  \\
0.6 & 4.42e-04 & 4.34e-04 & 6.23e-04 & 6.10e-04 & 4.28e-04  \\
0.7 & 6.05e-04 & 5.82e-04 & 8.26e-04 & 8.12e-04 & 5.79e-04  \\
0.8 & 7.71e-04 & 7.29e-04 & 1.04e-03 & 1.02e-03 & 7.29e-04  \\
0.9 & 9.57e-04 & 8.99e-04 & 1.27e-03 & 1.24e-03 & 8.99e-04  \\
1.0 & 1.15e-03 & 1.09e-03 & 1.51e-03 & 1.47e-03 & 1.10e-03  \\
\hline
\end{tabular}
\end{center}
\end{table*}

\section{Comparison with other work}
\subsection{Octree textures on a GPU}
\cite{Lefebvre_2005} have implemented an tree data structure for
a texture-mapping and tree traversal algorithm on a GPU. Owing to
the limitations of GPUs and on the SDK for the GPUs at that time, their
method seemed to be restricted to applications in computer
graphics. A critical point is that the possible depth of the tree
was limited, so that we cannot directly employ this implementation
for our purposes.

\subsection{Another tree implementation on a GPU}
\cite{Gaburov_2010} have reported another implementation of the
tree method on a GPU. Our implementation and their approach
share the same strategy, but there are differences in detail,
aside from the GPU architecture adopted. Both of us have
implemented a tree walk on a GPU. 
The implementation of \cite{Gaburov_2010} constructs interaction lists by means of a
tree walk on a GPU and then computes the force on the GPU using
these interaction lists. This implementation requires three
invocations of kernels. In contrast, we do a tree walk and compute
the force on-the-fly with one kernel invocation.

Both approaches have pros and cons. With the \cite{Gaburov_2010}
approach, a fairly high compute efficiency ($\sim$ 100 Gflops) has
been obtained, whereas our code shows low efficiency 
($\sim$ 30 Gflops). On the other hand, \cite{Gaburov_2010}'s code requires
more floating-point operations than does our optimal tree code.
We believe that our implementation is simpler than that of
\cite{Gaburov_2010}, which requires multi-pass computations. 
And we also believe that our implementation is easier to extend to a general
tree walk. In fact, we have extended our compute kernel written in
IL to the SPH method \citep{Gingold_1977, Lucy_1977} and obtained
fairly good performance.

\subsection{Fast multipole method on a GPU}
\cite{Gumerov_2008} have reported an implementation of a fast
multipole method (FMM) on a GPU. The FMM is a sophisticated
algorithm to evaluate long-range interaction forces with a
computational complexity of $O(N)$. In the FMM, in addition to the
replacement of distant particles with multipole expansions, local
expansions are utilized to evaluate the force acting on a group of
particles. \cite{Gumerov_2008} reported that for $N = 1,048,576$,
the algorithm took 0.68 s with $p = 4$ (Table 8 of
\cite{Gumerov_2008}), where $p$ is a parameter that controls the
error bound. Figure 10 of \cite{Gumerov_2008} indicates that the
average relative error obtained with $p = 4$ was $\sim$2$ \times
10^{-4}$ for the potential. The error is comparable to the
relative error obtained with the tree method with 
$\theta \sim$ 0.5--0.6. 
Note that Gumerov and Duraiswami used a random particle
distribution in a cube. For comparison, we did a test with a
similar particle distribution. Our kernel written in IL took 0.65
s for $N = 1,048,576$ with $\theta = 0.6$. 
The cache-hit rate of the test was 81\%. 
The performance of our tree code and that
obtained by Gumerov and Duraiswami with the FMM is comparable.
Note that the Cypress GPU used in the present work was different
from and newer than the GPU that Gumerov and Duraiswami used.
Generally, the FMM is well suited to applications that require
long-range interaction forces for uniformly distributed particles
or sources, whereas the tree method is more robust to the highly
clustered particles that typically arise in astrophysical
many-body simulations. 
We believe that our method is more suitable
than that of \cite{Gumerov_2008} for our purpose of astrophysical
applications.

\section{Conclusions}
In this paper, we have described our implementation of the tree
method on a Cypress GPU. By transforming a recursive tree
traversal into an iterative procedure, we have shown that the
execution of a tree traversal together with a force calculation on
a GPU can be practical and efficient. 
In addition, our implementation shows performance comparable to that of a recently
reported FMM code on a GPU.

We can expect to get further performance gains by fully utilizing
the four-vector SIMD operations of the SCs of the GPU. Moreover,
since 10--20\% of $T_{\rm total}$ is spent on the tree
construction, parallelization of this part on a multicore CPU will
be an effective way to boost the total performance. Provided that
we can easily extend our code to implement a force calculation for
short-range interactions by a method such as the SPH method, we
believe that a future extended version of our code will enable us
to do a realistic astrophysical simulation that involves baryon
physics with $N > 1,000,000$ very rapidly. It is fairly easy to
incorporate higher-order multipole expansion terms into our
method, and it would be a natural extension to the present work.
Another good application of the tree method on a GPU would be to
simulations that adopt a symmetrized Plummer potential \citep{Saitoh_2010}. 
We believe that our method is the best for
implementing that proposal, and hence that we shall certainly
obtain better accuracy and good performance in simulating galaxy
evolution and formation with different mass resolutions.

\section*{Acknowledgments}
The author would like to thank M.~Sato and K.~Fujiwara for their
efforts to utilize RV770 GPUs for astrophysical many-body particle
simulations. Part of this work is based on their undergraduate
theses of 2008 at the University of Aizu. The author is grateful
to Dr. H. Yahagi for his comments on a draft version of this
paper. This work was supported in part by a Grant-in-Aid from the
Ministry of Education (No. 21244020).

\appendix

\section{Working OpenCL code}
\label{app} 
In this appendix, we present our compute kernel written in C for OpenCL. 
We have tested this kernel with ATI Stream SDK 2.2. 
The function \verb|tree_gm| is an entry point to the kernel. 
\verb|n| is the number of particles. 
\verb|pos[i]| is a \verb|float4| variable for the positions and masses, i.e.,
$\mbox{\boldmath{$x$}}_i$ and $m_i$. 
\verb|acc_g[i]| is a \verb|float4| variable for the accelerations and the potential,
i.e., $\mbox{\boldmath{$a$}}_i$ and $p_i$. 
\verb|next[]| and \verb|more[]| are arrays that contains the two pointers described in Section \ref{NMp}. 
For all arrays, we assume that the data for the particles is at indices $i$ from $0$ to $n-1$. 
The data for the cells resides at $i >= n$. 
Finally, \verb|size[]| contains softening parameters of the particles for $i < n$ 
and the size of the cells for $i > = n$. 
We believe that it will be straightforward to prepare these arrays
from any tree implementation.

\begin{figure*}[t]
\centering
\begin{verbatim}
float4 g(float4 dx, float r2, float mj)
{
  float r1i = native_rsqrt(r2);
  float r2i = r1i*r1i;
  float r1im = mj*r1i;
  float r3im = r1im*r2i;
  float4 f;

  f.x = dx.x*r3im;
  f.y = dx.y*r3im;
  f.z = dx.z*r3im;
  f.w = r1im;

  return f;
}

__kernel void
tree_gm(__global float4 *pos,
        __global float *size,
        __global int *next,
        __global int *more,
        __global float4 *acc_g,
        int root, int n)
{
  unsigned int gid = get_global_id(0);
  float4 p = pos[gid];

  float4 acc = (float4)(0.0f, 0.0f, 0.0f, 0.0f);

  int cur = root;
  while(cur != -1) {
    float4 q = pos[cur];
    float  mj = q.w;
    float  s = size[cur];
    float4 dx = q - p;
    float  r2 = dx.x*dx.x + dx.y*dx.y + dx.z*dx.z;

    if (cur < n) {
      if (r2 != 0.0f) {
        r2 += s;
        acc += g(dx, r2, mj);
      }
      cur = next[cur];
    } else {
      if (s < r2) {
        acc += g(dx, r2, mj);
        cur = next[cur];
      } else {
        cur = more[cur];
      }
    }
  }

  acc_g[gid] = acc;
}
\end{verbatim}
\caption{
Working OpenCL code that that executes the iterative treewalk. 
The compute kernel that we have written in IL is mostly
similar to this code. 
}
\end{figure*}


\bibliographystyle{model5-names}
\bibliography{main}

\begin{thebibliography}{37}
\expandafter\ifx\csname natexlab\endcsname\relax\def\natexlab#1{#1}\fi
\providecommand{\bibinfo}[2]{#2}
\ifx\xfnm\relax \def\xfnm[#1]{\unskip,\space#1}\fi
\bibitem[{{AMD Inc.}(2010)}]{AMD_2010}
\bibinfo{author}{{AMD Inc.}} (\bibinfo{year}{2010}).
\newblock \bibinfo{title}{{ATI Stream Computing OpenCL Programming Guide,
  rev1.05}}.
\bibitem[{{Barnes}(1990)}]{Barnes_1990}
\bibinfo{author}{{Barnes}, J.} (\bibinfo{year}{1990}).
\newblock \bibinfo{title}{{A Modified Tree Code: Don't Laugh: It Runs}}.
\newblock {\it \bibinfo{journal}{Journal of Computational Physics}\/},  {\it
  \bibinfo{volume}{87}\/}, \bibinfo{pages}{161--170}.
\bibitem[{{Barnes} \& {Hut}(1986)}]{Barnes_1986}
\bibinfo{author}{{Barnes}, J.}, \& \bibinfo{author}{{Hut}, P.}
  (\bibinfo{year}{1986}).
\newblock \bibinfo{title}{{A Hierarchical $O(N \log N)$ Force-Calculation
  Algorithm}}.
\newblock {\it \bibinfo{journal}{Nature}\/},  {\it \bibinfo{volume}{324}\/},
  \bibinfo{pages}{446--449}.
\bibitem[{{Belleman} et~al.(2008){Belleman}, {B{\'e}dorf} \& {Portegies
  Zwart}}]{Belleman_2008}
\bibinfo{author}{{Belleman}, R.~G.}, \bibinfo{author}{{B{\'e}dorf}, J.}, \&
  \bibinfo{author}{{Portegies Zwart}, S.~F.} (\bibinfo{year}{2008}).
\newblock \bibinfo{title}{{High Performance Direct Gravitational $N$-body
  Simulations on Graphics Processing Units II: An Implementation in CUDA}}.
\newblock {\it \bibinfo{journal}{New Astronomy}\/},  {\it
  \bibinfo{volume}{13}\/}, \bibinfo{pages}{103--112}.
\bibitem[{{Bouchet} \& {Hernquist}(1988)}]{Bouchet_1988}
\bibinfo{author}{{Bouchet}, F.~R.}, \& \bibinfo{author}{{Hernquist}, L.}
  (\bibinfo{year}{1988}).
\newblock \bibinfo{title}{{Cosmological Simulations using the Hierarchical Tree
  Method}}.
\newblock {\it \bibinfo{journal}{Astrophysical Journal Supplement}\/},  {\it
  \bibinfo{volume}{68}\/}, \bibinfo{pages}{521--538}.
\bibitem[{{Dubinski}(1996)}]{Dubinski_1996}
\bibinfo{author}{{Dubinski}, J.} (\bibinfo{year}{1996}).
\newblock \bibinfo{title}{{A Parallel Tree Code}}.
\newblock {\it \bibinfo{journal}{New Astronomy}\/},  {\it
  \bibinfo{volume}{1}\/}, \bibinfo{pages}{133--147}.
\bibitem[{{Elsen} et~al.(2006){Elsen}, {Houston}, {Vishal}, {Darve}, {Hanrahan}
  \& {Pande}}]{Elsen_2006}
\bibinfo{author}{{Elsen}, E.}, \bibinfo{author}{{Houston}, M.},
  \bibinfo{author}{{Vishal}, V.}, \bibinfo{author}{{Darve}, E.},
  \bibinfo{author}{{Hanrahan}, P.}, \& \bibinfo{author}{{Pande}, V.}
  (\bibinfo{year}{2006}).
\newblock \bibinfo{title}{{$N$}-body simulation on {GPUs}}.
\newblock In {\it \bibinfo{booktitle}{SC '06: Proceedings of the 2006 ACM/IEEE
  Conference on Supercomputing}\/} (p. \bibinfo{pages}{188}).
\newblock \bibinfo{address}{New York}: \bibinfo{publisher}{ACM}.
\bibitem[{{Fujiwara} \& {Nakasato}(2009)}]{Fujiwara_2009}
\bibinfo{author}{{Fujiwara}, K.}, \& \bibinfo{author}{{Nakasato}, N.}
  (\bibinfo{year}{2009}).
\newblock \bibinfo{title}{{Fast Simulations of Gravitational Many-body Problem
  on RV770 GPU}}.
\newblock {\it \bibinfo{journal}{ArXiv e-prints}\/}, .
\bibitem[{{Fukushige} \& {Makino}(1996)}]{Fukushige_1996}
\bibinfo{author}{{Fukushige}, T.}, \& \bibinfo{author}{{Makino}, J.}
  (\bibinfo{year}{1996}).
\newblock In {\it \bibinfo{booktitle}{{Proceedings of Supercomputing '96}}\/}.
\newblock \bibinfo{address}{Los Alamitos}: \bibinfo{publisher}{IEEE Computer
  Society Press}.
\bibitem[{{Gaburov} et~al.(2010){Gaburov}, {B{\'e}dorf} \& {Portegies
  Zwart}}]{Gaburov_2010}
\bibinfo{author}{{Gaburov}, E.}, \bibinfo{author}{{B{\'e}dorf}, J.}, \&
  \bibinfo{author}{{Portegies Zwart}, S.~F.} (\bibinfo{year}{2010}).
\newblock \bibinfo{title}{{Gravitational Tree-code on Graphics Processing
  Units: Implementation in CUDA}}.
\newblock {\it \bibinfo{journal}{Procedia Computer Science}\/},  {\it
  \bibinfo{volume}{1}\/}, \bibinfo{pages}{1113--1121}.
\newblock \bibinfo{note}{ICCS 2010}.
\bibitem[{{Gingold} \& {Monaghan}(1977)}]{Gingold_1977}
\bibinfo{author}{{Gingold}, R.~A.}, \& \bibinfo{author}{{Monaghan}, J.~J.}
  (\bibinfo{year}{1977}).
\newblock \bibinfo{title}{{Smoothed Particle Hydrodynamics: Theory and
  Application to Non-spherical Stars}}.
\newblock {\it \bibinfo{journal}{Monthly Notices of the Royal Astronomical
  Society}\/},  {\it \bibinfo{volume}{181}\/}, \bibinfo{pages}{375--389}.
\bibitem[{{Greengard} \& {Rokhlin}(1987)}]{Greengard_1987}
\bibinfo{author}{{Greengard}, L.}, \& \bibinfo{author}{{Rokhlin}, V.}
  (\bibinfo{year}{1987}).
\newblock \bibinfo{title}{{A Fast Algorithm for Particle Simulations}}.
\newblock {\it \bibinfo{journal}{Journal of Computational Physics}\/},  {\it
  \bibinfo{volume}{73}\/}, \bibinfo{pages}{325--348}.
\bibitem[{{Gumerov} \& {Duraiswami}(2008)}]{Gumerov_2008}
\bibinfo{author}{{Gumerov}, N.~A.}, \& \bibinfo{author}{{Duraiswami}, R.}
  (\bibinfo{year}{2008}).
\newblock \bibinfo{title}{Fast multipole methods on graphics processors}.
\newblock {\it \bibinfo{journal}{Journal of Computational Physics}\/},  {\it
  \bibinfo{volume}{227}\/}, \bibinfo{pages}{8290--8313}.
\bibitem[{{Hamada} \& {Iitaka}(2007)}]{Hamada_2007}
\bibinfo{author}{{Hamada}, T.}, \& \bibinfo{author}{{Iitaka}, T.}
  (\bibinfo{year}{2007}).
\newblock \bibinfo{title}{{The Chamomile scheme: An Optimized Algorithm for
  $N$-body Simulations on Programmable Graphics Processing Units}}.
\newblock {\it \bibinfo{journal}{ArXiv Astrophysics e-prints}\/}, .
\bibitem[{{Hamada} et~al.(2009){Hamada}, {Narumi}, {Yokota}, {Yasuoka},
  {Nitadori} \& {Taiji}}]{Hamada_2009}
\bibinfo{author}{{Hamada}, T.}, \bibinfo{author}{{Narumi}, T.},
  \bibinfo{author}{{Yokota}, R.}, \bibinfo{author}{{Yasuoka}, K.},
  \bibinfo{author}{{Nitadori}, K.}, \& \bibinfo{author}{{Taiji}, M.}
  (\bibinfo{year}{2009}).
\newblock \bibinfo{title}{{42 TFlops Hierarchical $N$-body Simulations on GPUs
  with Applications in Both Astrophysics and Turbulence}}.
\newblock In {\it \bibinfo{booktitle}{SC '09: Proceedings of the Conference on
  High Performance Computing Networking, Storage and Analysis}\/} (pp.
  \bibinfo{pages}{1--12}).
\newblock \bibinfo{address}{New York}: \bibinfo{publisher}{ACM}.
\bibitem[{{Hernquist}(1990)}]{Hernquist_1990}
\bibinfo{author}{{Hernquist}, L.} (\bibinfo{year}{1990}).
\newblock \bibinfo{title}{Vectorization of tree traversals}.
\newblock {\it \bibinfo{journal}{Journal of Computational Physics}\/},  {\it
  \bibinfo{volume}{87}\/}, \bibinfo{pages}{137--147}.
\bibitem[{{Hockney} \& {Eastwood}(1981)}]{Hockney_1981}
\bibinfo{author}{{Hockney}, R.}, \& \bibinfo{author}{{Eastwood}, J.}
  (\bibinfo{year}{1981}).
\newblock {\it \bibinfo{title}{{Computer Simulation Using Particles}}\/}.
\newblock \bibinfo{address}{New York}: \bibinfo{publisher}{McGraw-Hill}.
\bibitem[{{Kawai} \& {Fukushige}(2006)}]{Kawai_2006}
\bibinfo{author}{{Kawai}, A.}, \& \bibinfo{author}{{Fukushige}, T.}
  (\bibinfo{year}{2006}).
\newblock \bibinfo{title}{{\$158/GFLOPS} astrophysical {$N$-body} simulation
  with reconfigurable add-in card and hierarchical tree algorithm}.
\newblock In {\it \bibinfo{booktitle}{SC '06: Proceedings of the 2006 ACM/IEEE
  Conference on Supercomputing}\/} (p.~\bibinfo{pages}{48}).
\newblock \bibinfo{address}{New York}: \bibinfo{publisher}{ACM}.
\bibitem[{{Kawai} et~al.(1999){Kawai}, {Fukushige} \& {Makino}}]{Kawai_1999}
\bibinfo{author}{{Kawai}, A.}, \bibinfo{author}{{Fukushige}, T.}, \&
  \bibinfo{author}{{Makino}, J.} (\bibinfo{year}{1999}).
\newblock \bibinfo{title}{{\$7.0/Mflops} astrophysical {$N$-body} simulation
  with treecode on {GRAPE-5}}.
\newblock In {\it \bibinfo{booktitle}{Supercomputing '99: Proceedings of the
  1999 ACM/IEEE Conference on Supercomputing (CDROM)}\/}
  (p.~\bibinfo{pages}{67}).
\newblock \bibinfo{address}{New York}: \bibinfo{publisher}{ACM}.
\bibitem[{{Kuijken} \& {Dubinski}(1995)}]{Kujiken_1995}
\bibinfo{author}{{Kuijken}, K.}, \& \bibinfo{author}{{Dubinski}, J.}
  (\bibinfo{year}{1995}).
\newblock \bibinfo{title}{Nearly self-consistent disc/bulge/halo models for
  galaxies}.
\newblock {\it \bibinfo{journal}{Monthly Notices of the Royal Astronomical
  Society}\/},  {\it \bibinfo{volume}{277}\/}, \bibinfo{pages}{1341--1353}.
\bibitem[{{Lefebvre} et~al.(2005){Lefebvre}, {Hornus} \&
  {Neyret}}]{Lefebvre_2005}
\bibinfo{author}{{Lefebvre}, S.}, \bibinfo{author}{{Hornus}, S.}, \&
  \bibinfo{author}{{Neyret}, F.} (\bibinfo{year}{2005}).
\newblock \bibinfo{title}{{Octree Textures on the GPU}}.
\newblock {\it \bibinfo{journal}{GPU Gems}\/},  {\it \bibinfo{volume}{2}\/},
  \bibinfo{pages}{595--614}.
\bibitem[{{Lucy}(1977)}]{Lucy_1977}
\bibinfo{author}{{Lucy}, L.~B.} (\bibinfo{year}{1977}).
\newblock \bibinfo{title}{{A Numerical Approach to the Testing of the Fission
  Hypothesis}}.
\newblock {\it \bibinfo{journal}{Astronomical Journal}\/},  {\it
  \bibinfo{volume}{82}\/}, \bibinfo{pages}{1013--1024}.
\bibitem[{{Makino}(1990)}]{Makino_1990}
\bibinfo{author}{{Makino}, J.} (\bibinfo{year}{1990}).
\newblock \bibinfo{title}{Vectorization of a treecode}.
\newblock {\it \bibinfo{journal}{Journal of Computational Physics}\/},  {\it
  \bibinfo{volume}{87}\/}, \bibinfo{pages}{148--160}.
\bibitem[{{Makino}(1991)}]{Makino_1991}
\bibinfo{author}{{Makino}, J.} (\bibinfo{year}{1991}).
\newblock \bibinfo{title}{Treecode with a special-purpose processor}.
\newblock {\it \bibinfo{journal}{Publications of the Astronomical Society of
  Japan}\/},  {\it \bibinfo{volume}{43}\/}, \bibinfo{pages}{621--638}.
\bibitem[{{Makino} \& {Taiji}(1998)}]{Makino_1998}
\bibinfo{author}{{Makino}, J.}, \& \bibinfo{author}{{Taiji}, M.}
  (\bibinfo{year}{1998}).
\newblock {\it \bibinfo{title}{{Scientific Simulations with Special-Purpose
  Computers: The GRAPE Systems}}\/}.
\newblock \bibinfo{address}{New York}: \bibinfo{publisher}{John Wiley and
  Sons}.
\bibitem[{{Nyland} et~al.(2007){Nyland}, {Harris} \& {Prins}}]{Nyland_2007}
\bibinfo{author}{{Nyland}, L.}, \bibinfo{author}{{Harris}, M.}, \&
  \bibinfo{author}{{Prins}, J.} (\bibinfo{year}{2007}).
\newblock \bibinfo{title}{{Fast $N$-body Simulation with CUDA}}.
\newblock In {\it \bibinfo{booktitle}{GPU Gems3}\/} (pp.
  \bibinfo{pages}{{677--696}}).
\newblock \bibinfo{address}{New York}: \bibinfo{publisher}{Addison-Wesley}.
\bibitem[{{Portegies Zwart} et~al.(2007){Portegies Zwart}, {Belleman} \&
  {Geldof}}]{Portegies_2007}
\bibinfo{author}{{Portegies Zwart}, S.~F.}, \bibinfo{author}{{Belleman},
  R.~G.}, \& \bibinfo{author}{{Geldof}, P.~M.} (\bibinfo{year}{2007}).
\newblock \bibinfo{title}{{High-performance Direct Gravitational $N$-body
  Simulations on Graphics Processing Units}}.
\newblock {\it \bibinfo{journal}{New Astronomy}\/},  {\it
  \bibinfo{volume}{12}\/}, \bibinfo{pages}{641--650}.
\bibitem[{{Saitoh} \& {Makino}(2010)}]{Saitoh_2010}
\bibinfo{author}{{Saitoh}, T.~R.}, \& \bibinfo{author}{{Makino}, J.}
  (\bibinfo{year}{2010}).
\newblock \bibinfo{title}{{The Natural Symmetrization for Plummer Potential}}.
\newblock {\it \bibinfo{journal}{ArXiv e-prints}\/}, .
\bibitem[{{Springel} et~al.(2005){Springel}, {White}, {Jenkins}, {Frenk},
  {Yoshida}, {Gao}, {Navarro}, {Thacker}, {Croton}, {Helly}, {Peacock}, {Cole},
  {Thomas}, {Couchman}, {Evrard}, {Colberg} \& {Pearce}}]{Springel_2005}
\bibinfo{author}{{Springel}, V.}, \bibinfo{author}{{White}, S.~D.~M.},
  \bibinfo{author}{{Jenkins}, A.}, \bibinfo{author}{{Frenk}, C.~S.},
  \bibinfo{author}{{Yoshida}, N.}, \bibinfo{author}{{Gao}, L.},
  \bibinfo{author}{{Navarro}, J.}, \bibinfo{author}{{Thacker}, R.},
  \bibinfo{author}{{Croton}, D.}, \bibinfo{author}{{Helly}, J.},
  \bibinfo{author}{{Peacock}, J.~A.}, \bibinfo{author}{{Cole}, S.},
  \bibinfo{author}{{Thomas}, P.}, \bibinfo{author}{{Couchman}, H.},
  \bibinfo{author}{{Evrard}, A.}, \bibinfo{author}{{Colberg}, J.}, \&
  \bibinfo{author}{{Pearce}, F.} (\bibinfo{year}{2005}).
\newblock \bibinfo{title}{{Simulations of the Formation, Evolution and
  Clustering of Galaxies and Quasars}}.
\newblock {\it \bibinfo{journal}{Nature}\/},  {\it \bibinfo{volume}{435}\/},
  \bibinfo{pages}{629--636}.
\bibitem[{{Sugimoto} et~al.(1990){Sugimoto}, {Chikada}, {Makino}, {Ito},
  {Ebisuzaki} \& {Umemura}}]{Sugimoto_1990}
\bibinfo{author}{{Sugimoto}, D.}, \bibinfo{author}{{Chikada}, Y.},
  \bibinfo{author}{{Makino}, J.}, \bibinfo{author}{{Ito}, T.},
  \bibinfo{author}{{Ebisuzaki}, T.}, \& \bibinfo{author}{{Umemura}, M.}
  (\bibinfo{year}{1990}).
\newblock \bibinfo{title}{{A Special-purpose Computer for Gravitational
  Many-body Problems}}.
\newblock {\it \bibinfo{journal}{Nature}\/},  {\it \bibinfo{volume}{345}\/},
  \bibinfo{pages}{33--35}.
\bibitem[{{Warren} et~al.(1998){Warren}, {Germann}, {Lomdahl}, {Beazley} \&
  {Salmon}}]{Warren_1998}
\bibinfo{author}{{Warren}, M.}, \bibinfo{author}{{Germann}, T.},
  \bibinfo{author}{{Lomdahl}, P.}, \bibinfo{author}{{Beazley}, D.}, \&
  \bibinfo{author}{{Salmon}, J.} (\bibinfo{year}{1998}).
\newblock \bibinfo{title}{{Avalon}: {An Alpha/Linux} cluster achieves 10
  {G}flops for \$15k}.
\newblock In {\it \bibinfo{booktitle}{Supercomputing '98: Proceedings of the
  1998 ACM/IEEE Conference on Supercomputing (CDROM)}\/} (pp.
  \bibinfo{pages}{1--11}).
\newblock \bibinfo{address}{Washington, DC}: \bibinfo{publisher}{IEEE Computer
  Society}.
\bibitem[{{Warren} \& {Salmon}(1992)}]{Warren_1992a}
\bibinfo{author}{{Warren}, M.}, \& \bibinfo{author}{{Salmon}, J.}
  (\bibinfo{year}{1992}).
\newblock \bibinfo{title}{Astrophysical {$N$-body} simulations using
  hierarchical tree data structures}.
\newblock In {\it \bibinfo{booktitle}{Supercomputing '92: Proceedings of the
  1992 ACM/IEEE Conference on Supercomputing}\/} (pp.
  \bibinfo{pages}{570--576}).
\newblock \bibinfo{address}{Los Alamitos, CA, USA}: \bibinfo{publisher}{IEEE
  Computer Society Press}.
\bibitem[{{Warren} \& {Salmon}(1993)}]{Warren_1993}
\bibinfo{author}{{Warren}, M.}, \& \bibinfo{author}{{Salmon}, J.}
  (\bibinfo{year}{1993}).
\newblock \bibinfo{title}{A parallel hashed oct-tree {$N$-body} algorithm}.
\newblock In {\it \bibinfo{booktitle}{Supercomputing '93: Proceedings of the
  1993 ACM/IEEE Conference on Supercomputing}\/} (pp. \bibinfo{pages}{12--21}).
\newblock \bibinfo{address}{New York}: \bibinfo{publisher}{ACM}.
\bibitem[{{Warren} et~al.(1997){Warren}, {Salmon}, {Becker}, {Goda}, {Sterling}
  \& {Wickelmans}}]{Warren_1997}
\bibinfo{author}{{Warren}, M.}, \bibinfo{author}{{Salmon}, J.},
  \bibinfo{author}{{Becker}, D.}, \bibinfo{author}{{Goda}, M.},
  \bibinfo{author}{{Sterling}, T.}, \& \bibinfo{author}{{Wickelmans}, G.}
  (\bibinfo{year}{1997}).
\newblock In {\it \bibinfo{booktitle}{{Proceedings of Supercomputing '97}}\/}.
\newblock \bibinfo{address}{Los Alamitos}: \bibinfo{publisher}{IEEE Computer
  Society Press}.
\bibitem[{{Warren} et~al.(1992){Warren}, {Quinn}, {Salmon} \&
  {Zurek}}]{Warren_1992}
\bibinfo{author}{{Warren}, M.~S.}, \bibinfo{author}{{Quinn}, P.~J.},
  \bibinfo{author}{{Salmon}, J.~K.}, \& \bibinfo{author}{{Zurek}, W.~H.}
  (\bibinfo{year}{1992}).
\newblock \bibinfo{title}{{Dark Halos Formed via Dissipationless Collapse. I:
  Shapes and Alignment of Angular Momentum}}.
\newblock {\it \bibinfo{journal}{Astrophysical Journal}\/},  {\it
  \bibinfo{volume}{399}\/}, \bibinfo{pages}{405--425}.
\bibitem[{{Warren} \& {Salmon}(1995)}]{Warren_1995}
\bibinfo{author}{{Warren}, M.~S.}, \& \bibinfo{author}{{Salmon}, J.~K.}
  (\bibinfo{year}{1995}).
\newblock \bibinfo{title}{{A Portable Parallel Particle Program}}.
\newblock {\it \bibinfo{journal}{Computer Physics Communications}\/},  {\it
  \bibinfo{volume}{87}\/}, \bibinfo{pages}{266--290}.
\bibitem[{{Yahagi} et~al.(1999){Yahagi}, {Mori} \& {Yoshii}}]{Yahagi_1999}
\bibinfo{author}{{Yahagi}, H.}, \bibinfo{author}{{Mori}, M.}, \&
  \bibinfo{author}{{Yoshii}, Y.} (\bibinfo{year}{1999}).
\newblock \bibinfo{title}{The forest method as a new parallel tree method with
  the sectional {V}oronoi tessellation}.
\newblock {\it \bibinfo{journal}{Astrophysical Journal Supplement}\/},  {\it
  \bibinfo{volume}{124}\/}, \bibinfo{pages}{1--9}.

\end{thebibliography}


\end{document}